\documentclass{IEEEcsmag}
\newcommand{\titlelong}{The Evolution of DNS Security and Privacy}

\usepackage[colorlinks,urlcolor=blue,linkcolor=blue,citecolor=blue,pdftex,
            pdfauthor={Levente Csikor, Dinil Mon Divakaran},
            pdftitle={\titlelong},
            pdfsubject={Submitted to IEEE Security \& Privacy Magazine},
            pdfkeywords={dns, privacy, security, encryption},
            pdfproducer={},
            pdfcreator={pdflatex}]{hyperref}
\expandafter\def\expandafter\UrlBreaks\expandafter{\UrlBreaks\do\/\do\*\do\-\do\~\do\'\do\"\do\-}
\usepackage{upmath,color}
\jvol{}
\jnum{}
\paper{}
\jmonth{Nov}
\jname{Submitted to IEEE Security \& Privacy Magazine (arXiv version)}
\jtitle{\titlelong}
\pubyear{2023}

\usepackage{amsmath}
\usepackage{comment}
\usepackage{tabularx}
\usepackage{caption}
\usepackage{subcaption}
\usepackage{booktabs}
\usepackage{float}
\usepackage{tikz}
\usepackage[normalem]{ulem}
\usepackage{graphicx}
\usepackage{adjustbox}
\usepackage{soul}
\usepackage{url}

\usepackage{tikz}

\newcommand{\rev}[1]{{\color{black} #1}}

\begin{document}

\title{\titlelong}

\author{{Levente Csikor}}
\affil{Institute for Infocomm Research (I$~^2$R), A*STAR, Singapore}

\author{Dinil Mon Divakaran}
\affil{National University of Singapore}

\markboth{REGULAR}{REGULAR}

\begin{abstract}
\rev{DNS, one of the fundamental protocols of the TCP/IP stack, has evolved over the years to protect against threats and attacks.
This study examines the risks associated with DNS and explores recent advancements 
that contribute towards making the DNS ecosystem resilient against various attacks while safeguarding user privacy.}
\end{abstract}

\maketitle


\chapteri{O}ver the past decade, the Internet privacy landscape has undergone a significant transformation, prompted by a widespread awareness of the need to safeguard personal data. 
This has led to the development of various security protocols and technologies.

\rev{The TCP/IP stack, that we rely on to communicate online, has made significant enhancements in terms of security and privacy.
For instance, Transport Layer Security (TLS) is now commonly used to secure applications such as web browsing (HTTPS), email transmissions (SMTPS, POP3S, IMAPS), and other services.}
IPSec and VPN enable secure communication at the network layer (IP layer), thereby protecting all application layer protocols with encryption during communications. 
\rev{Another tunneling method, The Onion Router (Tor) network and tools enable anonymous communication over the Internet by directing traffic through a volunteer overlay network. 
Unlike VPN, which tunnels user traffic through one (or two) secure end-points, ToR protects user privacy by encrypting and routing data through multiple volunteer-operated servers, making it challenging to trace the origin and destination of the communication.
Note, however, that tunneling methods invariably result in significantly slower connections and a compromised user experience.}

While individuals may perceive accessing websites via the HTTPS protocol (and seeing the green padlock next to their browser's address bar) as a secure connection, it is crucial to recognize that this alone does not guarantee \rev{complete protection}. 
Behind the scenes is a critical protocol and system, namely DNS (Domain Name System), that transparently aids in establishing connections.

\rev{Among other things (described below), the primary role of DNS is to translate user-friendly hostnames (like \texttt{example.com}) into machine-readable IP addresses (such as \texttt{93.184.216.34}).} 
Therefore, before connecting to any service, DNS resolution takes place as an initial step.
While DNS does not expose the actual content of any private information exchanged with a service (e.g., user credentials), the relatively slower adoption of robust security and privacy measures in this protocol opens up potential vulnerabilities. 
For example, today an adversary can interfere with a user's DNS traffic, and redirect them (via tampering of the resolved IP address) to an unintended website for malicious purposes (e.g., to steal credentials). 
Similarly, an intermediary network node having access to a user's DNS resolution process can profile user behavior based on the websites or domains they visit. 
The lack of adequate security measures in DNS resolution exposes users to risks that go beyond the perception of a secure HTTPS connection.

Recently, DNS underwent transformative security and privacy changes. 
\rev{This article presents its historical development,  current state, and challenges, and discusses potential future directions.}
Thus, we believe this article contributes to comprehending the DNS protocol, its security measures, and broader online safety considerations. 

\section{BACKGROUND: DNS ECOSYSTEM}
\label{sec:dns_ecosystem}
\begin{figure*}[h]
     \centering
     \includegraphics[width=.95\textwidth]{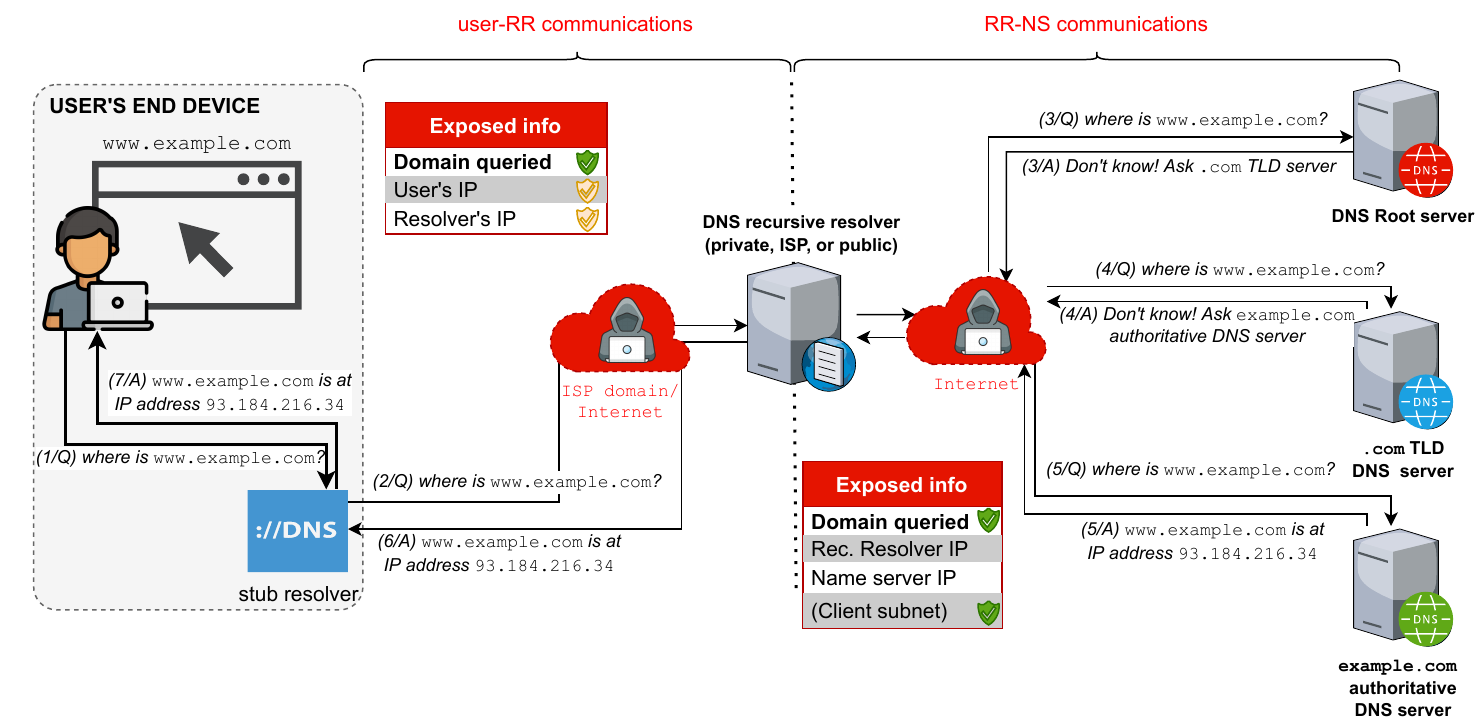}
\caption{The DNS machinery. 
The tables ``Exposed info'' during the user-RR and RR-NS communications illustrate the DNS data that can potentially be exposed to malicious intermediary nodes intercepting the communication. Shield icons indicate the types of information that can be safeguarded through encryption (green) or make them difficult to reveal (yellow).}
\label{fig:dns_ecosystem}
\end{figure*}

\rev{One of the main roles of the Domain Name System (DNS)
is to facilitate the translation of domain names to machine-understandable IP addresses.}
When functioning properly, DNS operates transparently, often unnoticed by regular users and even technical administrators. 
This seemingly unobtrusive nature may lead to underestimating the significance of DNS, assuming it requires only minimal safeguarding, and that its security is adequately addressed by other application defenses (e.g., encrypted web/email communications). 

\rev{Before delving into the security and privacy intricacies of DNS, 
let us begin with a basic example to grasp the fundamental address translation service of DNS.}
Figure~\ref{fig:dns_ecosystem} illustrates a scenario where a user intends to visit a website, say \texttt{www.example.com}. 
To establish a connection, the domain name of the website must first be translated into an IP address through the DNS resolution process.

First, the browser consults the local DNS stub resolver, which caches domain-IP mappings (step 1/Q).
As there is no existing entry, the stub resolver seeks assistance from a DNS recursive resolver (step 2/Q).
A DNS recursive resolver (RR) retrieves the requested information by communicating with name servers.
The RRs are typically operated by ISPs, for their customers; but they can also be publicly available services maintained by Internet giants like Cloudflare, Google, or Quad9. 
Note, the RR is \textit{not} responsible for any specific domain as it is not an {\em authoritative name server} for the domains. 
It efficiently resolves DNS queries by navigating the DNS hierarchy on behalf of the client.
There is a significant proliferation of RRs across the internet, with their numbers surpassing six million~\cite{dns_rr_number}. 

Like the client's stub resolver, an RR maintains a cache for quicker DNS resolution. 
However, an RR serves an entire subnet (i.e., ISP network) or a larger portion of Internet users in the case of public recursive resolvers (e.g., Google's 8.8.8.8), leading to a much larger cache compared to stub resolvers. 
Consequently, for the majority of users accessing popular domains, the DNS resolution often ends here, as RR can promptly respond from its cache. 
This stage is denoted via the \textit{"user-RR communications"} in Figure~\ref{fig:dns_ecosystem}.

When there is no cached entry for a specific domain, the RR initiates a DNS resolution process. 
This process involves multiple queries and responses until the particular IP address of the requested domain is obtained. 
Initially, the RR queries one of the 13 root name servers (step 3/Q in Figure~\ref{fig:dns_ecosystem}). 
These root name servers serve as the initial point for resolving domain names from scratch, providing information about the authoritative name servers responsible for top-level domains (TLDs) like \texttt{.com}, \texttt{.org}, and country-code TLDs such as \texttt{.uk} or \texttt{.jp} (denoted by step 3/A in our example). 
While the root servers do not directly handle queries for \rev{fully qualified domain names (FQDN)}, their role is essential in the overall functioning of the DNS system. 

Subsequently, the RR consults the TLD name server (e.g., the \texttt{.com} TLD name server - step 4/Q in Figure~\ref{fig:dns_ecosystem}. 
Similar to the root name server, the TLD DNS server is not responsible for the FQDN either; it directs (step 4/A) the RR to the authoritative name server (ANS) in step 5/Q.
\rev{ANS is a DNS server responsible for providing authoritative information about an FQDN, i.e., about a specific (sub)domain.}
It holds the official up-to-date records (e.g., IP addresses) of the domain, allowing it to respond to DNS queries for that domain with authoritative answers.
In our illustration, the \texttt{example.com} name server holds authority over its subdomain (i.e., \texttt{www.example.com}).
Consequently, it provides the correct IP address of the server hosting the intended web content (step~5/A).

Once the RR obtains the accurate domain-to-IP address mapping, it responds to the user's stub resolver (step 6/A), which in turn responds to the user's browser. 
At the same time, both the stub resolver and the RR temporarily cache the information locally. 
Cached entries serve quick responses for further requests, but expire mandates to repeat the process, ensuring fresh data\footnote{Unlike CPU or routing caches, DNS cached entries are not ``kept warm'' after hits. 
Validity relies on the domain's ANS, with preset expiration (minutes to days) based on factors like availability, load balancing, and security\footnote{\rev{\url{https://tinyurl.com/38xwhdwp}}}.}.

\rev{
Beyond domain name to IP address mapping, DNS includes various other resource record types. 
For instance, the PTR (Pointer) record is used for reverse DNS lookups (i.e., for IP-to-domain name mapping). The TXT  record stores arbitrary text data and is often utilized for domain validation process for issuing TLS certificates. 
For instance, the Let's Encrypt certificate authority requests the domain owner to create a specific TXT record containing a predefined value in the DNS configuration. 
By successfully adding this TXT record to the DNS, the domain owner demonstrates control over the domain, thus validating ownership.}

\section{SECURITY AND PRIVACY OF PLAIN-TEXT DNS}
\label{sec:privacy_and_security_of_dns}

\subsection{Security: Attacks on DNS}
\label{sec:privacy_and_security_of_dns__security}

DNS protocol typically operates in plaintext, exposing exchanges between clients and servers (any hierarchy level) to eavesdropping and tampering. 
Malicious actors include malevolent ISPs, governments, compromised routers, and interceptors at open mobile hotspots.
The plain-text DNS is vulnerable to various types of attacks, including spoofing, cache poisoning, DNS tunneling, DNS dangling, and DNS amplification attacks. 

\subsubsection{\rev{DNS cache poisoning:}} 
\label{sec:privacy_and_security_of_dns__security__spoofing}

\rev{DNS cache poisoning, also known as DNS spoofing, is an attack in which the DNS cache of a recursive resolver is manipulated to divert traffic to malicious websites or compromise user privacy.
Attackers can either be on-path (e.g., Monster-in-the-middle) or off-path. 
In the former, the attacker is positioned on the network path between the DNS resolver and the authoritative DNS server, and intercepts and manipulates DNS query and response packets.
In an off-path attack, however, the attacker does not intercept the communications directly but attempts to guess the transaction IDs used in DNS queries, successfully injecting forged responses into the resolver's cache.
To impede rapid guessing of the correct transaction ID, UDP source ports were introduced for verification, elevating the entropy from 65,536 to approximately 134 million guesses. However, it's important to note that side channels present in Linux kernels for over a decade were disclosed only a couple of years ago~\cite{dns_cache_poisoning_side_channel}.

DNS Security Extensions (DNSSEC) addresses DNS cache poisoning by digitally signing DNS data, thereby ensuring its integrity. 
It establishes a chain of trust through cryptographic keys, allowing DNS resolvers to validate signatures and reject tampered or unauthorized DNS information. 
DNSSEC provides end-to-end security by safeguarding the entire DNS resolution path from the root DNS server to authoritative name servers. 
This prevents attackers from successfully injecting malicious data into DNS resolvers' caches, enhancing the overall security of the Domain Name System. 
Despite its potential benefits and being a standard since 1999, DNSSEC has not seen widespread adoption. 
\rev{As reported by APNIC\footnote{\url{https://stats.labs.apnic.net/dnssec}}, 
the global DNSSEC validation rate is only $\sim30$\%.}
Major drawbacks include the complexity of implementation, maintenance, and potentially frequent outages if not applied properly\footnote{\rev{IANIX, ``Major DNSSEC Outages and Validation Failures'' \url{https://tinyurl.com/3nyexkjs}}}. 
Additionally, DNSSEC introduces a significant overhead in terms of computational resources and an increase in response sizes, which can impact DNS performance.
}

\subsubsection{DNS amplification:}
\label{sec:privacy_and_security_of_dns__security__amplification}

In this attack, the attacker spoofs the source IP address and sends, typically by means of a botnet, a large number of DNS queries to open DNS recursive resolvers. 
These RRs, unaware of the spoofing, respond to the queries by sending responses to the victim's IP address (spoofed in the DNS query packets by the attacker), overwhelming their network capacity and causing a denial-of-service at the victim.
For the amplification effect, an attacker crafts DNS queries that have a small request size but elicit an order of magnitude larger response sizes from the DNS resolvers.
\rev{It has been shown recently that the amplification factor of DNS-based amplification attacks can even be as high as 98~\cite{dns_amplification_microsoft}. }

Note that malicious actors have the capability to launch these attacks at any stage of the DNS resolution process.
The impact and effectiveness vary by the stage at which they occur. 
For example, if an attacker successfully poisons the cache of an RR, all users relying on that resolver will be affected. 
On the other hand, if only a single user's stub resolver is compromised,
only that particular user will be impacted.

From a defense perspective, RRs may have specific countermeasures (e.g., logs, history to detect anomalies) in place to 
be more resilient against these threats. 
In contrast, a user accessing the Internet via open WiFi hotspots, is typically accepting DNS responses without validation, thus becoming more vulnerable for attackers. 
Later, we discuss countermeasures that such users can employ while using WiFi hotspots.

\subsection{Privacy Implications}
\label{sec:privacy_and_security_of_dns__privacy}
DNS privacy pertains to safeguarding, or the lack thereof, user-sensitive information and browsing activities during the DNS resolution process. 
While the exposure of privacy through DNS may not appear as malicious as stealing credentials, it undeniably risks exposing one of the most revealing aspects of our online presence: our Internet browsing patterns.

\rev{The exposure of our DNS information transforms it into a valuable asset not just for malicious actors aiming to profile user behavior, but also for purposes such as targeted advertisements, distributing malicious files, conducting phishing attacks, or redirecting to other forms of suspicious or malicious activities.}
DNS communications can be exploited for nationwide surveillance (e.g., for understanding political preferences, socio-religious standing). 
States might also leverage DNS to track online activities of certain interest groups (e.g., extremists). 

In Figure~\ref{fig:dns_ecosystem}, we divided the resolution process into two stages: user-RR (i.e., before recursive resolution) and the RR-NS communications (after recursive resolution).
This separation effectively illustrates the extent of exposure of specific privacy-sensitive user information at different stages of the DNS hierarchy. 
It is important to note that in the following discussion, we assume standard DNS, meaning DNS data is transmitted in plain text, and hence can be monitored. 
However, the choice between using the default connection-less UDP protocol or the more reliable connection-oriented TCP protocol (which is a default fallback mechanism if a UDP response is too big and hence would need to be fragmented) does not affect the significance of the exposed information as TCP does not provide encryption either.

\subsubsection{User-RR communications:}
\label{sec:privacy_and_security_of_dns__privacy__before}
The information that is exposed through plain-text DNS message exchanges in the user-RR communications includes \textit{(i)}~the domain name itself, \textit{(ii)}~the IP address of the requesting user, and \textit{(iii)}~the IP address of the resolver that the user is configured to use.

Undoubtedly, the most privacy-sensitive information among the exposed data is the combination of the requested domain name and the identity (IP address) of the user making the request. 
In particular, this information can be used to identify and track a user's online activities. 
It allows entities with access to this information to create user profiles, monitor browsing behavior, and potentially link it to other identifiable information.
The exposed DNS information can be  further utilized for profiling user's interests, preferences, and online behavior, which can then be exploited for purposes such as targeted advertisements .
Governments, ISPs, or other entities can exploit this information to carry out surveillance for law enforcement purposes, tracking of individual's internet use, and censor and filter contents.
Last but not least, aggregating DNS information from multiple sources can provide insights into broader trends, patterns, and user behaviors. This data can be used for various purposes, such as market research, trend analysis, or understanding Internet use patterns on a larger scale.

\rev{In contrast, \textit{(iii)} only reveals the IP address of the RR used, providing a limited value for an eavesdropper.
However, an attacker may glean some information from this, such as the user's level of familiarity with different resolver options.}
For instance, if the user is using a publicly available resolver instead of one provided by their ISP, it could suggest a deliberate choice made by the user. 
This phenomenon has been documented in various cases, motivated by factors such as performance improvements or circumventing DNS-based censorship efforts by ISPs or governments\footnote{\rev{\url{https://tinyurl.com/y4xacx5m}}}. 
The new DNS protocols discussed below (e.g., DoT/DoH), shield the sensitive information from eavesdroppers.

\subsubsection{RR-NS communications:}
\label{sec:privacy_and_security_of_dns__privacy__after}

\rev{As depicted in Figure~\ref{fig:dns_ecosystem}, the exposure of user information is minimized following recursive resolution. Therefore, to compromise user privacy and unveil user identity, it is advantageous for an adversary to intervene in DNS traffic during user-RR communications.}
In RR-NS communications, DNS packets lack user IP, reducing direct tracking. 
Only the requested domain name is accessible.

\rev{Following the recent QNAME minimization standard~\cite{qname_min}, the user's request undergoes a process where only the pertinent parts are transmitted to the corresponding entities in the DNS hierarchy. 
This minimizes the exposure of the requested domain and reduces overall overhead.
In our example, when the RR receives the user's request for \texttt{www.example.com}, only \texttt{.com} is sent to the root server, followed by transmitting only \texttt{example.com} to the TLD DNS server. 
The full \texttt{www.example.com} is relayed to the authoritative server (ANS) at the final stage only.}

Compared to the prior stage, as DNS traffic reaches the ANS, the resolver's IP becomes that of the RR. 
However, RRs might still disclose the user's subnet through DNS Client Subnet~\cite{dns_client_subnet}, sharing a part of the client's IP (subnet) with ANS. 
This extension offers more accurate answers based on the client's network location, being a trade-off between performance enhancement and heightened privacy risk.

Like the prior phase, if RR enforces encryption, DNS data remains secure. 
\rev{Yet, RRs lack standardized encryption methods\footnote{\rev{Note, DNSCurve is specifically designed for securing the communication between the recursive resolver and the authoritative nameserver, however, it is not standardized by the IETF body.}} and motivation to safeguard communication at this stage.}
Each encryption layer increases response time, risking user migration to faster alternatives.

Protocol weaknesses expose varying attacker capabilities. 
User-RR communications can be intercepted by any entity at hotspots, while RR-NS comms demand potent adversaries like ISPs or a state-sponsored actor.

\section{EVOLUTION OF DNS}
\label{sec:dns_privacy_evolution}
Next, we examine the important measures developed in an attempt to resolve the above-mentioned issues. 
As we present the recent DNS encryption protocols, we discuss their functionalities and their differences.
In the subsequent sections of the paper, we shift our focus to security measures implemented for user-RR communications, as this is the area where regular Internet users have the ability to safeguard their own privacy.
Furthermore, there is no standardized method to encrypt RR-NS communications.

\subsection{Secure DNS Extensions}
\label{sec:dns_privacy_evolution__protocols}

\subsubsection{DNSCrypt:}
\label{sec:dns_privacy_evolution__protocols__dnscrypt}
DNSCrypt\footnote{\url{https://www.dnscrypt.org/}} is a cryptographic protocol to enhance the security and privacy of user-RR communications.
Unlike DNSSEC, the protocol encrypts the DNS traffic
between a client and a DNS recursive resolver, preventing eavesdropping and tampering. 
It ensures the integrity and confidentiality of DNS queries and responses. 
DNSCrypt operates at the transport layer either over UDP or TCP, and by default, it runs over port 443.

DNSCrypt debuted in 2011 and is used by OpenDNS. 
Yet, the lack of standardization by the IETF has hindered its widespread adoption. 
Lately, DNSCrypt has been growing steadily and many popular DNS RR providers (e.g., Cloudflare,  Quad9, OpenDNS) support DNSCrypt.
However, the focus on encrypted DNS has shifted towards two standardized encryption protocols instead: DNS-over-HTTPS (DoH) and DNS-over-TLS (DoT). 
These protocols have gained traction by providing a standardized and widely supported approach to offer enhanced security and privacy for user-RR communications.

\subsubsection{DNS-over-TLS:}
\label{sec:dns_privacy_evolution__protocols__dot}
In 2015, the IETF introduced DNS-over-TLS (DoT,~\cite{dot}) as a standard to encrypt user-RR DNS communications. 
DoT utilizes the widely adopted Transport Layer Security (TLS) protocol to encrypt and authenticate the DNS traffic. 
Thereby, it provides a secure and private channel for user-RR communication.
DoT (DNS-over-TLS) uses port 853 instead of the commonly used ports \texttt{53} (plain-text DNS) or \texttt{443} (HTTPS),  primarily to ensure that DoT traffic is distinguishable from regular unencrypted DNS traffic. 
This allows administrators to apply rules to DoT traffic for proper routing and handling.

Due to its message format similarity with plain-text DNS, adopting DoT as the secure alternative is relatively straightforward.
DoT enables easy implementation at an operating system-wide level, allowing all applications on the same system to benefit from its security. 
Currently, DoT is supported by all major OSes, including mobile OSes.

\subsubsection{DNS-over-HTTPS:}
\label{sec:dns_privacy_evolution__protocols__doh}
In 2017, the IETF introduced DNS-over-HTTPS (DoH)~\cite{doh}, which offers the same level of encryption as DoT but utilizes the well-known port number \texttt{443}---port 443 is used for encrypted web communications (HTTPS). 
Consequently, DoH, like DoT and DNSCrypt, ensures that the user-RR communications 
remain unreadable to eavesdroppers. 
On the other hand, DoH operates in the application layer and was initially developed as an experimental extension for Firefox browsers. 
This enabled a faster time-to-market since its adoption did not depend on underlying operating system support. 
Within a year of its standardization in 2018, Mozilla introduced full support for Trusted Recursive Resolvers (TRR) in its browser, and Google made similar changes to Chrome in 2019. 
Additionally, in 2020, Mozilla announced that it would use DoH, specifically Cloudflare's TRR, as the default option for US users. 
As DoH can be easily enabled in browsers, its widespread adoption among users is expected to surpass other alternatives. 
Furthermore, major (mobile) operating systems now also have native implementations of DoH. 
Currently, there are around 450 publicly available DoH resolvers\footnote{\url{https://github.com/curl/curl/wiki/DNS-over-HTTPS}}, with this number anticipated to grow substantially in the wild.

\subsubsection{DNS-over-QUIC:}
\label{sec:dns_privacy_evolution__protocols__doq}
DNS-over-QUIC (DoQ) is a protocol that uses QUIC (Quick UDP Internet Connections) for user-RR DNS communications.
QUIC is a recent UDP-based transport protocol developed by Google to overcome the limitations of traditional TCP and also align well with the demands and capabilities of 5G networks. 
Unlike TCP's three-way handshake, QUIC employs a zero-round-trip time (0-RTT) handshake, significantly reducing connection setup time. 
QUIC also supports multiplexing, allowing multiple data streams to be sent simultaneously over a single connection, which eliminates the head-of-line blocking issue of TCP. 
Additionally, QUIC incorporates built-in encryption using TLS~1.3, ensuring secure data transmission by safeguarding against eavesdropping and preserving data privacy.

DoQ takes advantage of QUIC's benefits and applies them to DNS resolution. 
By encapsulating DNS communications within QUIC packets, DoQ improves the performance and security of DNS communications.
Like DoT, DoQ uses port \texttt{853}.
Any DNS resolver supporting QUIC can deploy DoQ.
However, the penetration of DoQ worldwide is still in its infancy. 
Research studies, such as~\cite{doq_first_look}, have identified over 1,000 DoQ resolvers in the wild. 
However, when it comes to the publicly available ones, their count remains in the single digits, AdGuard\footnote{\url{https://adguard-dns.io/kb/general/dns-providers/}} being the first to deploy one in 2020.

\subsubsection{DNS-over-HTTPS3:}
\label{sec:dns_privacy_evolution__protocols__doh3}
\rev{HTTP/3, the third major iteration of the Hypertext Transfer Protocol (HTTP), succeeds HTTP/2, aiming to enhance both performance and security. Employing QUIC as its transport layer, HTTP/3 introduces DNS-over-HTTP/3 (DoH3), a novel protocol combining the advantages of DNS-over-HTTPS (DoH) with the performance enhancements of HTTP/3. DoH3 utilizes the same encryption as DoH while incorporating the HTTP/3 protocol.
DoH3 is to HTTP3 what DoH is to HTTP2, utilizing port \texttt{443}. Unlike DoQ, DoH3 employs HTTP-based messages and operates at the application layer. The primary distinction lies in the fact that DoH3 requires DNS resolvers to concurrently support QUIC and HTTP/3.}

\subsubsection{Decentralized and Oblivious DNS:}
\label{sec:dns_privacy_evolution__protocols__odns}

In the pursuit of privacy enhancements through encrypted DNS and public recursive resolvers, users inevitably encounter an important consideration: while their DNS data remains protected from eavesdroppers, it is ultimately exposed to the recursive resolver \textit{per se}. 
This raises the fundamental question of whether it is preferable to share DNS data with a third-party company that may operate in a different and distant country with varying legal jurisdictions rather than with their local Internet Service Provider (ISP). 
Besides, as more users reconfigure their systems to use the same publicly available RRs, the DNS ecosystem becomes more centralized, with a few large companies getting access to most DNS resolutions. 
This means users have to carefully assess the level of trust they place in different entities when handling their sensitive DNS data. 
To address this concern, further extensions have been developed.

\begin{figure}[h!]
    \centering
    \includegraphics[width=.95\linewidth]{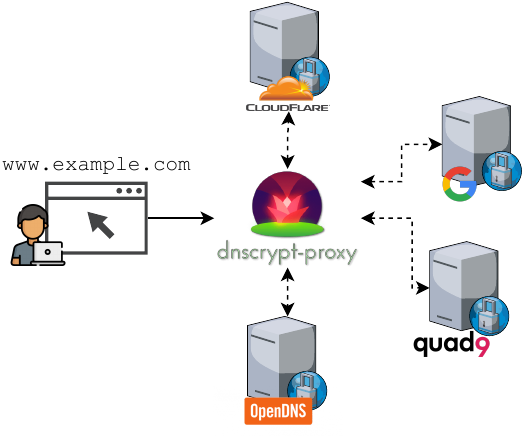}
    \caption{DNSCrypt sends different queries to different resolvers to avoid any of them having a full profile of the client.}
    \label{fig:dnscrypt-proxy}
\end{figure}

\textit{Decentralized DNS} involves an architectural extension that distributes the client's DNS queries to multiple resolvers in a sophisticated manner (see Figure~\ref{fig:dnscrypt-proxy}). 
For example, DNScrypt-proxy\footnote{\url{https://github.com/DNSCrypt/dnscrypt-proxy}} supports DNScrypt, DoH, and uses hashing (by default) to route queries to resolvers.
Unlike a simple round-robin approach, hashing ensures that the same set of domains will consistently be resolved by the same resolvers, thereby preventing any single resolver from having a complete profile of the user. 
This strategy helps to alleviate the privacy concerns associated with centralized DNS resolution and further enhances the privacy of DNS queries.

\begin{figure}[!h]
    \centering
    \includegraphics[width=.95\linewidth]{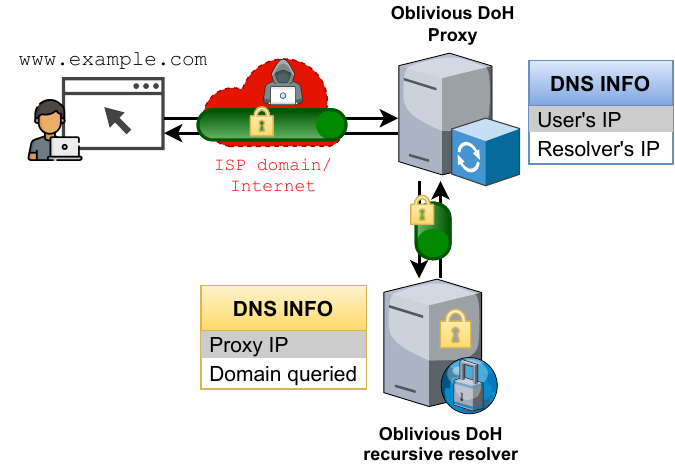}
    \caption{Oblivious DNS. The user's query is encrypted with the ODoH RR; hence the proxy only knows the user's IP, while the ODoH RR only knows the queried domain.}
    \label{fig:odoh}
\end{figure}

\textit{Oblivious DNS} is a set of solutions that provide enhanced security transparently within the protocol level.  
Solutions such as Anonymized DNSCrypt and Oblivious DoH (ODoH) share a common principle (cf.~Figure~\ref{fig:odoh}).
Instead of directly connecting to a Recursive Resolver (RR), clients first connect to a transparent proxy. 
This proxy acts as an intermediary and relays the DNS queries to the actual RR, effectively concealing the clients' real IP addresses.  
Additionally, the user's DNS query is encrypted using the public key of the RR; hence the relaying proxy only knows the IP address but cannot decrypt the actual DNS query. 
On the other hand, the RR will only know the DNS query (after decrypting it) with the clients' IP addresses appearing as the relay's IP.

To be effective, it is crucial that the proxy and the target resolver are operated by separate entities and do not collude. 
As long as there is no collusion between the proxy and the RR, an attacker can only succeed if both the proxy and the target are compromised.
\rev{At the same time, this approach also resolves the above-mentioned issue around who a user might want to share its DNS data with; neither a third-party company nor its local ISP.}

Although ODoH~\cite{odoh} is a relatively new protocol with an experimental standard status at IETF,
some major DNS resolvers, such as Cloudflare and NextDNS, already support the protocol.

\section{\rev{CENSORSHIP ATTACKS}}
\label{sec:censorship}
\subsection{\rev{DNS Practices}}
On the one hand, the plain-text DNS data is useful for numerous valuable services such as parental control, malware, and malicious domain detection, or even captive portals that redirect to access control pages. 
On the other hand, DNS is also often misused. 
In addition to monetizing and user profiling discussed earlier, 
DNS can be exploited for censorship 
by ISPs (mostly for law enforcement) and authoritarian regimes.
For instance, DNS-based blocking is implemented in multiple countries (e.g., China, Iran) to restrict political speech and occasionally during periods of political unrest in various regions across the world~\cite{pets_foci_censorship}. 
These instances demonstrate DNS's role in censorship, emphasizing the urgency for privacy safeguards and preventing misuse of DNS. These are known as DNS censorship attacks.

We elaborate on to what extent the above-mentioned encrypted DNS solutions are able to counter DNS censorship attacks.
Consider the scenario where a user seeks to circumvent ISP or country-wide censorship by using encrypted DNS and configuring its system to use a remote RR. 
Although this method secures DNS from interception, its encrypted nature may arouse censorship suspicions, leading to potential communication blockage.

\begin{figure}[h!]
    \centering
    \includegraphics[width=.95\linewidth]{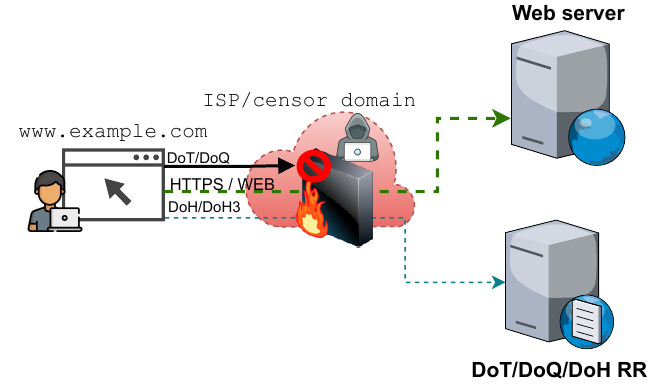}
    \caption{DoT/DoQ can be easily filtered and downgraded to plain-text DNS by censors, but DoH and DoH3 blend into regular HTTPS web traffic, making them indistinguishable and difficult to block.}
    \label{fig:dns_censoring}
\end{figure}

In particular, if the user employs DoT/DoQ on the well-known dedicated port, the communication becomes easily distinguishable and filterable. 
As depicted in Figure~\ref{fig:dns_censoring}, censors could potentially block all TCP (for DoT) and UDP (for DoQ) packets destined to any remote service on port \texttt{853}. 
This filtering's impact is clear as key services (e.g., browsing, email, plain-text DNS) use distinct ports, reducing risk of collateral damage.

Note, filtering DoT/DoQ communications forces the user (e.g., browser/OS) to downgrade to plain-text DNS, which might occur seamlessly due to the {\em opportunistic privacy profiles} that most stub resolvers adopt. 
Once the downgrade happens, the censor can continue employing traditional DNS censoring mechanisms (for domains of interest) without hindrance.

When users use (oblivious) DoH or DoH3, utilizing HTTPS on port 443, encrypted DNS merges with regular HTTPS traffic, becoming indistinguishable and hard to filter (see Figure~\ref{fig:dns_censoring}). 
While censors may attempt to use more sophisticated methods, including machine learning models for identification of DoH~\cite{doh_ml}, such a model's capability to identify the protocol comes with a non-zero false positive rate that results in wrong traffic being affected; besides, clients can use packet padding techniques~\cite{doh_ml} to counter such models.

\subsection{\rev{Beyond DNS Practices and Countermeasures}}
\rev{Although encrypting DNS and subsequent communication (i.e., HTTPS) boost user privacy, certain information pertaining to a connection can still expose information that can be used for profiling \textit{and} censorship.
The compromise of user privacy is possible if an adversary has an extensive and up-to-date set of sites and their IP addresses~\cite{what_can_you_learn_from_IP}; however maintaining such data is challenging, and the use of virtualization, public clouds, and content delivery networks makes it more complex to associate specific users with online activities.
HTTPS sites require TLS certificates for encrypted communication and identity verification, but the Server Name Indication (SNI) field in the plain-text TLS handshake, indicating the desired certificate for the visited domain (e.g., \texttt{www.example.com}), could also be exploited for user profiling and censorship attacks~\cite{pets_foci_censorship}.}

\rev{To address these challenges, various enhancements have been developed and are currently undergoing further development.
For instance, Apple's recently developed privacy feature, {\em Private Relay}, functions akin to oblivious DNS but for web browsing. It encrypts user Internet traffic, passing it through two relays of two parties. 

To conceal the SNI field, TLS 1.3 introduced Encrypted SNI, which encrypts the SNI using the server's public key distributed through DNS. 
Although eSNI represents a substantial advancement, it does not meet the objective of attaining complete handshake encryption and has brought forth additional challenges\footnote{Cloudflare blogpost: \url{https://tinyurl.com/5cvjwjmc}}.
}

\subsubsection{Encrypted Client Hello (ECH):}
ECH is an improvement over eSNI introduced in TLS 1.3, offering a more generic and flexible solution. 
ECH encrypts the entire \texttt{Client Hello} message, \rev{not only} the SNI field, using a special encryption scheme.
The public keys for encryption are also distributed via DNS; however, there are several differences \rev{compared to eSNI}. 
First, the encryption and the key management are usually done via a third party (i.e., not by the administrator of the actual service the user intends to connect to \rev{like in the case of eSNI}), also known as a ``client-facing server'' --- typically a CDN.
Second, the public key for encryption is stored in the DNS \texttt{HTTPS} resource record of the CDN domain \rev{not as a TXT record of the actual service's DNS data}. 
It also means that when the client resolves the domain name of a service, the corresponding records (i.e., IP addresses) will point to the client-facing server.
Third, in the event that the public key obtained via the DNS \texttt{HTTPS} record expires or becomes compromised, a mechanism is in place to ensure enhanced reliability, \rev{while eSNI would simply fail}.

\rev{
\section{KEY TAKEAWAYS AND FUTURE DIRECTIONS}
Starting as a plain-text protocol, DNS has evolved significantly to having multiple protocols that strengthen its security posture. 
The development of DoT, DoH, DoQ, and ODoH protocols offers options for average users to encrypt and safeguard their communications. 
And there is a noticeable uptick in the adoption of these protocols~\cite{adoption_encrypted_dns}.
While not impervious to determined and powerful adversaries, this shift is still significant as it enables a user to maintain communication privacy, particularly in regions where regulatory entities do not obstruct such communications.

In regions with heightened censorship measures, as previously discussed, the sole reliance on these protocols does not ensure the desired level of privacy. 
Additional measures like ECH become imperative to shield connections from potential metadata leaks of SNI. 
It is crucial to recognize, however, that such enhancements are ineffective in securing connection attempts to encrypted DNS resolvers, creating a pragmatic loop when relying on DNS for acquiring encryption keys. 
In censoring web connections relying on ECH, an authoritarian censor might surveil preceding encrypted DNS connections based on their SNI information. 
Upon identification, it can obstruct these connection attempts, leading to a scenario where user connections are either disrupted or revert to plain-text DNS, thus resetting the privacy progress.

However, on a positive note, researchers are working to address this concern; for instance, recent works \cite{blindTLS,quicstep} propose to bypass TLS SNI censorship through a QUIC/TCP-TLS handshakes conducted within an encrypted tunnel first (e.g., via an encrypted proxy), concealing the SNI field from censors, then seamlessly migrating the session. 
While they are intended to be alternatives to ECH, they might be used for tunneling the initial plain-text handshake with an encrypted DNS resolver, then migrate the session; note encrypted DNS sessions are long-lived, hence doing this is required once only when initializing the connection.
We believe this evolution converges to a point where average users are able to hide their DNS communications completely without depending on a separate tool such as VPN or Tor.

}

\def\refname{REFERENCES}
\bibliographystyle{ieeetr}
\bibliography{main.bib}

\begin{IEEEbiography}{Levente Csikor}
(csikor\_levente@i2r.a-star.edu.sg) is a research scientist at the Institute for Infocomm Research (I$^2$R), A*STAR, Singapore. 
His main scope of research focuses on future internet architectures, DNS privacy, and network security.
He obtained his MSc. and Ph.D. degrees from the Budapest University of Technology and Economics, Hungary.
\end{IEEEbiography}

\begin{IEEEbiography}{Dinil Mon Divakaran}
(Senior Member, IEEE; dinil@comp.nus.edu.sg) is an Adjunct Assistant Professor of the School of Computing, National University of Singapore (NUS). 
His research spans topics such as phishing, network attacks, security log analysis, DNS, and IoT security.
\end{IEEEbiography}

\end{document}